\documentclass[epj]{svjour}
\usepackage{graphics}
\begin{document}
\title{Study of the sextic and decatic anharmonic oscillators using an interpolating scale function}
\author{K. Manimegalai\inst{1} \and Swaraj Paul\inst{2} \and M. M. Panja\inst{3} \and Tapas Sil\inst{1}    
}                     
\offprints{\email{tapassil@iiitdm.ac.in}}          
\institute{Department of Physics, Indian Institute of Information Technology Design and Manufacturing Kancheepuram, Chennai-600127, Tamil Nadu, India  \and Discipline of Mathematics, Indian Institute of Technology Indore, Simrol, Indore 453 552,  India \and Department of Mathematics, Visva-Bharati, Santiniketan 731235, West Bengal, India 	
}
\date{Received: date / Revised version: date}
%
\abstract{
	 Anharmonic oscillators with the sextic and decatic potentials are studied employing  the refinable interpolating scale functions. This method  yields  highly accurate values of both energy eigenvalues and eigenfunctions  for the sextic and decatic oscillator without constraining the potential parameters. Convergence of the solutions in the present method is noticed to be very fast.
%
} 
\maketitle
\section{Introduction}
The anharmonic oscillator (AHO) is a system generalizing the simple linear harmonic oscillator and is widely used for modeling many physical phenomena for understanding them, especially systems involving quantum mechanical vibration \cite{maiz2018sextic,gaudreau2015computing,bonham1966use,bender1968analytic,bender1969anharmonic,biswas1973eigenvalues,pathak2002classical,gomez2000bound,bera2012homotopy,patnaik1990anharmonic,chaudhuri1991improved}.
{
	There are only a few numbers of quantum systems exist, e.g., hydrogen atom and harmonic oscillator which have exact solutions. It is difficult to find solutions for the problems involving anharmonicity, and hence, one adopt either a numerical or an analytical approximation method to get the energy eigenvalues and eigenfunctions of such systems. The solutions by means of any new method are usually validated by comparing the energy eigenvalues and eigenfunctions with the available exact values or those obtained from different approximation methods. AHOs also serve as a test bench for different approximate methods of getting solutions of Schr$\ddot{\textrm{o}}$dinger equation and to examine the ability of an approximation method to reproduce the dependence of the eigenvalues on the coefficients of the potential \cite{bender1993analytic}. Most of the AHOs belongs to the family of quasi-exactly solvable potentials for which the exact energy eigenvalues and eigenfunctions can be obtained for a few low lying states  \cite{banerjee1978k,turbiner1988av,adhikari1989,ushveridze2017quasi,bera2007generalization}. Attempts are going on for getting solutions for higher states or for unrestricted values of the parameters admissible to the existence of  bound states for such potentials  \cite{flessas1983non,voros1994exact,bay1997spectrum,lay1997quartic}. 
	
	Recently, a renewed interest has been noticed for getting analytical solutions in closed form for the Schr$\ddot{\textrm{o}}$dinger equation with the sextic, and decatic anharmonic potentials which find wide application in different branches of physics \cite{somorjai1962double,wall1937double,buganu2017shape,quigg1979quantum}. It is shown in Ref. \cite{maiz2018sextic,brandon2013exact}, that under certain conditions on the parameters of the potential, the energy eigenvalues and eigenfunctions can be found exactly in terms of the potential parameters. Gaudreau {\it et. al.} \cite{gaudreau2015computing}, on the other hand, employed the Sinc Collocation Method (SCM) with double exponential transformation and computed energy eigenvalues numerically for the arbitrary values of the potential parameters with desired accuracy.
	
	Theory of wavelet finds a lot of applications in various fields of science and engineering,
	\cite{kessler2004,van2004wavelets,farge2015wavelet,panja2016,paul2016multiscale,paul2018use}
	especially it is found to be very effective for getting  very accurate numerical solutions of  differential and integral equations. 
    Recently, interpolating wavelet method \cite{deslauriers1989symmetric,anasari1991,donoho1992interpolating,saito1993multiresolution,bertoluzza1996wavelet,holmstrom1999solving,shi2001}  has attracted the attention of the researchers as it has an added 
    advantage that the coefficients of wavelet can be calculated from direct combinations of discrete  samples rather than the inner product integrals. This method is found to be very efficient and yields not only very accurate results but also the convergence of the solutions are very fast \cite{lippert1998,dyn2002}.  We intend to use refinable interpolating scale function (IPSF) approach 
    for studying the aforementioned AHOs.

	This paper is organized as follows. In section 2, the definition and some basic properties of IPFS relevant for the present study is discussed. Applications of  the method based on IPSF to the quantum mechanical AHOs have been discussed in section 3. Finally, we conclude the paper in section 4 with some discussion on the implications of the findings.
	
	\section{Formulation}
	Interpolating scaling function was introduced by Deslauries and Dubuc \cite{deslauriers1989symmetric} using the symmetric iterative interpolation based on Lagrange interpolation. We are going to use the Deslauries-Dubuc interpolating scaling function (DDIPSF) to evaluate the energy eigenvalues and the eigenfunctions of the Schr$\ddot{\textrm{o}}$dinger equation with potentials of our interest. The DDIPSF, $\Phi(x)$ follows the basic properties as given below:
	\begin{itemize}
		\item[(i)] $\Phi(x)$ is a compactly supported scaling function supported on $[{-N+1},{N-1}]$ \cite{donoho1992interpolating,saito1993multiresolution}.
		\item[(ii)]  $\int_{-\infty}^{\infty}\Phi(x)dx=1=\sum\limits_{k=-\infty}^{\infty}\Phi(x-k).$
		\item[(iii)]  $\Phi(x)$ of order $N$ ($N$ must be positive even integer)  interpolates the Kronecker sequence at the integers $\lbrace -\frac{N}{2}+1,..,\frac{N}{2}\rbrace$.
		\item[(iv)] The two scale relation or refinement equation among the $\Phi(x)$'s is,
		\begin{equation}\label{tsr}
		\Phi(x)=\sum_{k=-N+1}^{N-1}a_k\Phi(2x-k),
		\end{equation}
		where, $a_i=\Phi(\frac{i}{2})$, and $a_{1-2i}=l_i(\frac{1}{2})$, where $l_i(x)$ is the Lagrange polynomial of order $i$,\ $i=\lbrace -\frac{N}{2}+1,..,\frac{N}{2}\rbrace$ \cite{deslauriers1989symmetric}.
		For an illustration, the values of the coefficients $a_k$ are given in Table 1 for $N=4$. As, $a_k=a_{-k}$, the values of $a_k$ are displayed in the table only for $k=0$ and for the allowed negative values of $k$.
	
		\begin{table}\label{akvalues}
			\begin{center}
			\caption{The Values of $k$ and $a_k(=a_{-k})$ for N=4}
				\begin{tabular}{|c|c|c|c|c|}
					\hline
					$k$ & $-3$ & $-2$ & $-1$ & $0$   \\
					\hline
					$a_k$ & $-\frac{1}{16}$ & $0$ & $\frac{9}{16}$ & $1$   \\
					\hline				
				\end{tabular}
			\end{center}
		\end{table}
	
		\item[(v)] The DDIPSF follows the reflection symmetry, $\Phi(x)=\Phi(-x)$.
		\item[(vi)] For $j\ge 0$, $\Phi_{j,k}(x)=2^{\frac{j}{2}}\Phi(2^jx-k)$.
		\item[(vii)]  For any $f$, being a polynomial of degree $(N-1)$, the coefficients $\beta_{j,k}$  in the representation, $f=\sum\limits_k\beta_{j,k}\Phi_{j,k}$ can be recovered by sampling $\beta_{j,k}=\frac{1}{2^{\frac{j}{2}}} f(2^{-j}k)$    \cite{donoho1992interpolating}.
	\end{itemize} 
	The time independent Schr$\ddot{\textrm{o}}$dinger equation in one dimension is written as ($\hbar=2m=1$),
	\begin{equation}\label{tise}
	-\frac{d^2\psi(x)}{dx^2}+(V(x)-E)\psi(x)=0,\ \ \ \ \ \ x\in {\bf R}.
	\end{equation}
	As the wavefunction $\psi(x) \in L^2(\bf R) $, it can be expanded in terms of $\Phi_{j,k_1}(x)$ at resolution $j\ (j\in {\bf N})$ and with the translation $k_1$ as,
	\begin{equation}\label{basisexp}
	\psi(x)=\sum\limits_{k_1\in \wedge_{j}}c_{j,k_1}\Phi_{j,k_1}(x),
	\end{equation}
	where, $c_{j,k_1}$ is the coefficients and $\wedge_{j}$ is the appropriate index set $\lbrace k_{\textrm{min}},k_{\textrm{min}}+1,...,k_{\textrm{max}}\rbrace$. We are interested to solve eq.(\ref{tise}) on the domain $[-6,6]$ then $k_{\textrm{min}}=\{-6(2^{j})+N-1\}$ and $k_{\textrm{max}}=\{6(2^{j})-N+1\}$. 
	Substituting the wavefunctions $\psi(x)$ from eq.(\ref{basisexp}) to eq.(\ref{tise}), we multiply $\Phi_{j,k_2}(x)$ on the both sides of the equation and then integrate with respect to $x$ from $-\infty$ to $\infty$ which yields,
	\begin{eqnarray}\label{sle}
	-\sum\limits_{k_1\in \wedge_j}\left\langle\frac{d^2\Phi_{j,k_1}}{dx^2},\Phi_{j,k_2}\right\rangle c_{j,k_1}+\sum\limits_{k_1\in \wedge_j}\left\langle V(x)\Phi_{j,k_1},\Phi_{j,k_2}\right\rangle c_{j,k_1}
	=E\sum\limits_{k_1\in \wedge_j}\left\langle \Phi_{j,k_1},\Phi_{j,k_2}\right\rangle c_{j,k_1},\ \ \ \ \ \ k_2\in \wedge_j,
	\end{eqnarray}
	where, $\langle x,y \rangle$ is the inner product defined by,
	\begin{equation}
	\left\langle x,y \right\rangle=\int_{-\infty}^{\infty}x(t)y(t)dt.
	\end{equation}
	Here. the eq.(\ref{sle}) may be regarded as a matrix eigenvalue problem, and can be written as,
	\begin{equation}\label{matrixevp}
	A_j\tilde{c_j}=EB_j\tilde{c_j},
	\end{equation}
	where, the matrix elements ${A_j}_{N(\wedge_j)\times N(\wedge_j)}\ \left(=(a_{j,k_1,k_2})_{k_1,k_2\in\wedge_j}\right)$  and ${B_j}_{N(\wedge_j)\times N(\wedge_j)}\ \left(=(b_{j,k_1,k_2})_{k_1,k_2\in\wedge_j}\right)$ are given by,
	\begin{equation}
	a_{j,k_1,k_2}=-{\bf L}_{j,k_1,k_2}+{\bf I}_{j,k_1,k_2}.
	\end{equation}
	and,
	\begin{equation}
	b_{j,k_1,k_2}=\langle\Phi_{j,k_1},\Phi_{j,k_2}\rangle,
	\end{equation}
	respectively.
	 Matrices, ${\bf L}_{j,k_1,k_2}$ and ${\bf I}_{j,k_1,k_2}$ are given by,
	 \begin{equation}
	 \label{lmatrix}
	 {\bf L}_{j,k_1,k_2}=\left\langle\frac{d^2\Phi_{j,k_1}}{dx^2},\Phi_{j,k_2}\right\rangle,
	 \end{equation}
	 and,
	 \begin{equation}
     {\bf I}_{j,k_1,k_2}=\left\langle V(x)\Phi_{j,k_1},\Phi_{j,k_2}\right\rangle.	
     \label{imatrix} 
     \end{equation} 

	\subsection{Evaluation of ${\bf L}_{j,k_1,k_2}$}
	Let us use the property (vi) of $\Phi_{j,k_i}\ (i=1,2)$ at higher resolution $j$ and we get from eq.(\ref{lmatrix}),
	\begin{equation}
	{\bf L}_{j,k_1,k_2}
	=2^{2j}{\bf L}_{0,0,k_2-k_1}=
	0\ \ \textrm{when}\ |k_2-k_1|\geq 2N-2.
	\end{equation}
	Now, using the two scale relation of $\Phi$ mentioned in (iv), we get a system of homogeneous equations for ${\bf L}_k(={\bf L}_{0,0,k})$ and can be written as,
	\begin{equation}
	{\bf L}_k=2\sum\limits_{l_1=-N+1}^{N-1}\sum\limits_{l_2=-N+1}^{N-1}a_{l_1}a_{l_2}{\bf L}_{2k+l_2-l_1}.
	\end{equation}
	Hence, in order to get a non-trivial solution of ${\bf L}_k$, we need to construct a nonhomogeneous equation as follows,
	\begin{equation}\label{nh}
	x^2=\sum\limits_{k\in{\bf Z}}k^2\ \Phi(x-k).
	\end{equation}
	Taking double derivative on the both sides of eq.(\ref{nh}) with respect to $x$, we multiply $\Phi(x)$ throughout the equation and then integrating  from $x=-\infty$ to $x=\infty$, we get, 
	\begin{equation}
	\sum\limits_{k=-N+1}^{N-1}k^2\ {\bf L}_k=2.
	\end{equation}
	The value of ${\bf L}_k$ is zero when $|k|\geq 2N-2$. For $N=4$, the values of ${\bf L}_k$ where  $|k|< 2N-2$ are given in Table 2.	
	\begin{table}[ht]\label{Lkvalues}
		\begin{center}
		\caption{The Values of ${\bf L}_k(={\bf L}_{-k})$ when $|k|\leq 5$}
			\begin{tabular}{|c|c|c|c|c|c|c|}
				\hline
				$k$ & $-5$ & $-4$& $-3$ & $-2$ & $-1$ & $0$   \\ \hline
				${\bf L}_k$ & $0$& $0$& $-\frac{1}{72}$ & $0$ & $\frac{9}{8}$ & $-\frac{20}{9}$   \\ \hline
			\end{tabular}
				\end{center}
	\end{table}
	\subsection{Evaluation of ${\bf I}_{j,k_1,k_2}$ for a given form of potential $V(x)=x^{m}$}
	Let us rewrite ${\bf I}_{j,k_1,k_2}$ in eq.(\ref{imatrix}) as,
	$${\bf I}_{j,k_1,k_2}=\langle x^m\Phi_{j,k_1},\Phi_{j,k_2}\rangle={\bf H}_{j,m,k_1,k_2}\ (\textrm{say}).$$
	Using the property (vi) of $\Phi$ at resolution $j$, we get,
	\begin{equation}
	{\bf H}_{j,m,k_1,k_2}=\frac{1}{2^{jm}}{\bf H}_{m,k_1,k_2},
	\end{equation}
	with, 
	\begin{eqnarray}\label{hmk1k2}
	{\bf H}_{m,k_1,k_2}=\left\lbrace\begin{array}{ll}
	{\bf H}_{m,k_1} & k_2=0\\
	\sum\limits_{r=0}^{m}{m \choose r} 
	k_2^{m-r}{\bf H}_{r,k_1-k_2} & k_2\neq 0\\
	0 & |k_1-k_2|\geq 2N-2.
	\end{array}
	\right..
	\end{eqnarray}
	To obtain the values of ${\bf H}_{m,k_1,k_2}$, we use the two scale relation for $\Phi$.
	Now, ${\bf H}_{m,k}$ is expressed as,
	\begin{eqnarray}\label{hmk}
	{\bf H}_{m,k}&=&\left \langle x^m\Phi(x-k),\Phi(x)\right\rangle\nonumber\\
	    &=& \left\langle x^m\sum\limits_{l_1=-N+1}^{N-1}a_{l_1}\Phi(2x-2k-l_1),\sum\limits_{l_2=-N+1}^{N-1}a_{l_2}\Phi(2x-l_2)\right\rangle.
	\end{eqnarray} 
	Using the expression of ${\bf H}_{m,k}$ in eq.(\ref{hmk1k2}) with $k_1=0$, and $k_2=k$, we get the  recurrence relation from eq.(\ref{hmk}) as given below, 
	\begin{eqnarray}
	{\bf H}_{m,k}=\frac{1}{2^{m+1}}\left\lbrace\sum\limits_{l_1=-N+1}^{N-1}\sum\limits_{l_2=-N+1}^{N-1}a_{l_1}a_{l_2} \left({\bf H}_{m,2k+l_2-l_1}
	 +\sum\limits_{s=0}^{r-1}{r \choose s} 
	l_1^{r-s}{\bf H}_{s,2k+l_2-l_1}\right)\right\rbrace.
	\end{eqnarray}
	This recurrence relation is used to find out the values ${\bf H}_{m,k}$. 
	The value of ${\bf H}_{m,k}$ is zero when $|k|\geq 2N-2$. The values of ${\bf H}_{m,k}$ are given in Table 3 for $|k|< 2N-2$ with $N=4$.
		\subsection{Evaluation of $b_{j,k_1,k_2}$}
	We can easily evaluate the value of $b_{j,k_1,k_2}$ as, 
	\begin{eqnarray}
		b_{j,k_1,k_2}&=&\langle \Phi_{j,k_1},\Phi_{j,k_2}\rangle\nonumber \\
		&=&\langle x^0\Phi_{j,k_1},\Phi_{j,k_2}\rangle \nonumber \\
		&=&{\bf H}_{j,0,k_1,k_2}.{\small {\tiny }}
	\end{eqnarray}
	Using the procedure of computing ${\bf H}_{j,0,k_1,k_2}$, one can calculate the values of $b_{j,k_1,k_2}$.
	\begin{table}\label{Hvalues}
		\begin{center}
		\caption{The Values of ${\bf H}_{m,k}(={\bf H}_{m,-k})$ for $m=0,2,4,6,8,10$ when $|k|\leq 5$. $u(-n)$ indicates $u \times 10^{-n}$.}
				\begin{tabular}{|c|c|c|c|c|c|c|}
				\hline
				$k$ & $-5$ & $-4$ & $-3$ & $-2$ & $-1$ & $0$   \\
				\hline
				${\bf H}_{0,k}$ & $-1.48291(-7)$ & $-7.59247(-5)$ & $2.79513(-3)$ & $-4.02449(-2)$ & $1.37042(-1)$ & $8.00968(-1)$   \\
				\hline
				${\bf H}_{2,k}$ & $-9.29064(-7)$ & $-3.08303(-4)$ & $6.68093(-3)$ & $-5.1135(-2)$ & $-6.32431(-4)$ & $9.07914(-2)$   \\
				\hline
				${\bf H}_{4,k}$ & $-5.87691(-6)$ & $-1.32529(-3)$ & $1.96084(-2)$ & $-1.07352(-1)$ & $-8.28606(-2)$ & $4.38699(-2)$ \\
				\hline
				${\bf H}_{6,k}$ & $-3.75209(-5)$ & $-5.96315(-3)$ & $6.73162(-2)$ & $-2.26355(-1)$ & $-2.21724(-1)$ & $5.92416(-2)$   \\
				\hline
				${\bf H}_{8,k}$ & $-2.41636(-4)$ & $-2.76491(-2)$ & $2.60004(-1)$ & $-3.55121(-1)$ & $-5.59101(-1)$ & $1.28922(-1)$   \\
				\hline
				${\bf H}_{10,k}$& $-1.56844(-3)$ & $-1.29915(-1)$ & $1.08057$ & $3.09366(-1)$ & $-1.51288$ & $3.32909(-1)$   \\
				\hline
			\end{tabular}
         \end{center}
	\end{table}
\section{Results and discussions}
In this section, we present, the energy eigenvalues and eigenfunctions for AHOs using DDIPSF scheme. The explicit form of $\Phi$ is not known, however, their values at dyadic points are obtained by using properties (ii) and (iii). Diagonalizing the matrix $A_j$ in eq.(\ref{matrixevp}), we get the energy eigenvalues of the AHO. Corresponding to a particular eigenvalue, we evaluate $\tilde{c_j}$ which in turn gives us the energy eigenfunction. We have considered the order of the scaling functions $\Phi$ as $N=4$  to get reasonable accuracy with optimal computational time. The calculation is done for $x$ values between 
$-6$ to $6$. 

\subsection{Sextic potential}
The sextic AHO with even parity is represented by the potential,
\begin{equation}\label{poten6}
V(x)=ax^2+bx^4+cx^6.
\end{equation} 
The sextic potentials play an important role in studying the spectra of molecules such as ammonia and hydrogen bonded solids, nuclear shape, and to model for quark confinement in quantum chromodynamics \cite{somorjai1962double,wall1937double,buganu2017shape,quigg1979quantum}.
A lot of studies have been done on the sextic anharmonic potential to test newly developed calculational scheme. In most of the cases, the comparison of energy eigenvalues and eigenfunctions are done with the existing exact values. Recently, Maiz \textit{et al.}\cite{maiz2018sextic} have given the exact energy eigenvalues and eigenfunctions for the ground state as well as a few excited states with some constraints on the potential parameters ($a,b,\ \textrm{and} \ c$) in eq.(\ref{poten6}). In Table 4, we present the energy eigenvalues of the ground state and few excited states using the method based on DDIPSF  and compare them with those obtained by  Maiz \textit{et al.} \cite{maiz2018sextic}. Since the potential parameters are interrelated according to the prescription of Maiz {\it et. al.}, for a particular state, there is only a combination of parameters for which they can give exact energy eigenvalue.  We have presented the energy eigenvalues calculated from the scheme based on DDIPSF for $j=3,5$ and $7$ in the fourth, fifth and sixth column, respectively. One may find that as one goes for  a higher resolution, i.e., for a  larger value of $j$, the maximum deviation of energy eigenvalue from the exact one is going to a reduced value ($\approx 10^{-9}$ for $j=5$ and  $\approx 10^{-12}$ for $j=7$). It may be noted that in all cases  shown in the table, the maximum deviation in our results from the corresponding exact values is  negligible ($\approx 10^{-12}$) which indicates the high efficiency of our method to produce very accurate energy values. Moreover, we are able to calculate the energy eigenvalues for the values of the parameters of the potential other than the constraints values i.e. for arbitrary values of the parameters and displayed in Table 4. 
\begin{table}[ht]
	\begin{center}
	\caption{Comparison of energy eigenvalues ($E_n$) with the available exact values in Ref.\cite{maiz2018sextic} with $b=c=1$ for $j=3,5,7.$}\label{maiz6}
		\begin{tabular}{|c|c|c|c|c|c|}
			\hline
			$E_n$ & $a=$  & From  & DDIPSF  & DDIPSF & DDIPSF \\
			&$\frac{b^2}{4c}-\alpha \sqrt{c}$  & \cite{maiz2018sextic} & for $j=3$ &  for $j=5$ &  for $j=7$\\
			\hline
			$E_0$ & $\alpha=3$ & $0.5$ & $0.500003$ & $0.5000000008$ & $0.4999999999949$  \\
			& $5$ &   & $-0.344956$ & $-0.3449627092$ & $-0.3449627110455$  \\
			& $7$ &   & $-0.359511$ & $-0.3595439975$ & $-0.35954400614$  \\
			\hline
			$E_1$ & $\alpha=3$ &   & $3.120048$ & $3.1200310533$ & $3.1200310489652$  \\
			& $5$ & ${1.5}$ & $1.500022$ & $1.5000000058$ & $1.5000000000001$  \\
			& $7$ &   & $-1.499985$ & $-1.4999999959$ & $-1.4999999999839$  \\
			\hline
			$E_2$ & $\alpha=3$ &   & $7.942999$ & $7.9429050475$ & $7.9429050229149$  \\
			& $5$ &  & $6.197359$ & $6.1972676365$ & $6.1972676124053$  \\
			& $7$ & $4.5$ & $4.500098$ & $4.5000000261$ & $4.5000000000210$  \\
			\hline
		\end{tabular}
	\end{center}
\end{table}
It is obvious that the DDIPSF scheme using higher resolution ($j$) gives rise to more accurate energy eigenvalue. But calculation with large value of $j$ requires large computing time $t_j$. We note the computation time $t_j$ for various $j$ (using a computer with processor, i7 and RAM=8GB) for the ground state ($E_0=0.5$) of the sextic potential taking $b=c=1,$ and $\alpha=3$ which read as  $t_3=0.58s$, $t_5=8.86s$,  and $t_7=135.33s$. This indicates that the calculation of energy in DDIPSF is quite fast for $j\le 7$ whereas the accuracy achieved for this $j (=7)$  is very high ($\approx 10^{-12}$). 

\begin{figure}
	\centering
	\begin{minipage}{.4\textwidth}
		\resizebox{0.85\textwidth}{!}{\includegraphics{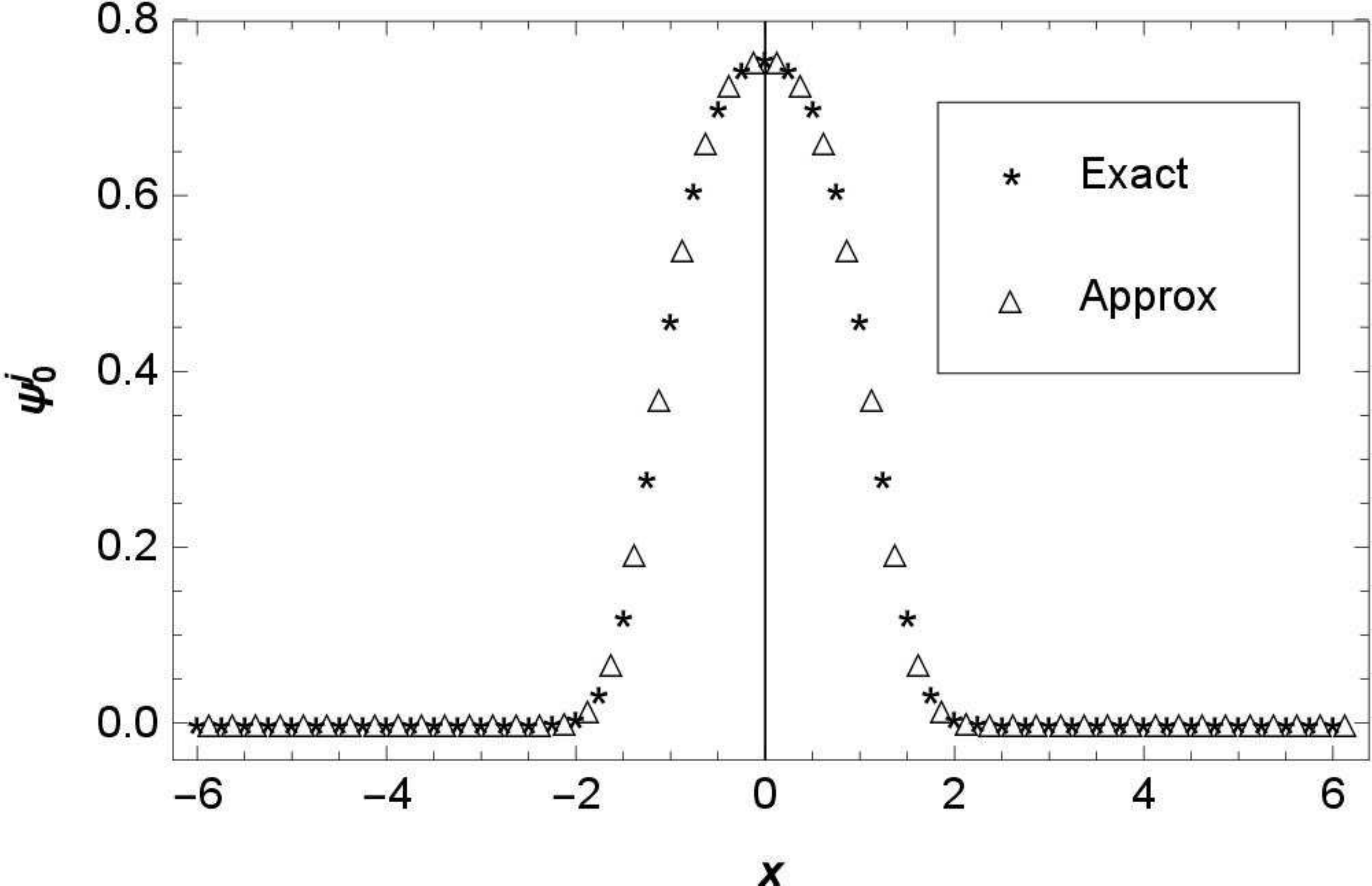}}
		\caption{ Comparison of normalize ground state eigen function ($\Psi_0$) with the exact one \cite{maiz2018sextic} for the potential eq.(\ref{poten6}) with $a=-\frac{11}{4},\ b=1,\ c=1$.}
		\label{fig1}
	\end{minipage}\qquad
	\bigskip
	\begin{minipage}{.4\textwidth} 
		\resizebox{0.85\textwidth}{!}{\includegraphics{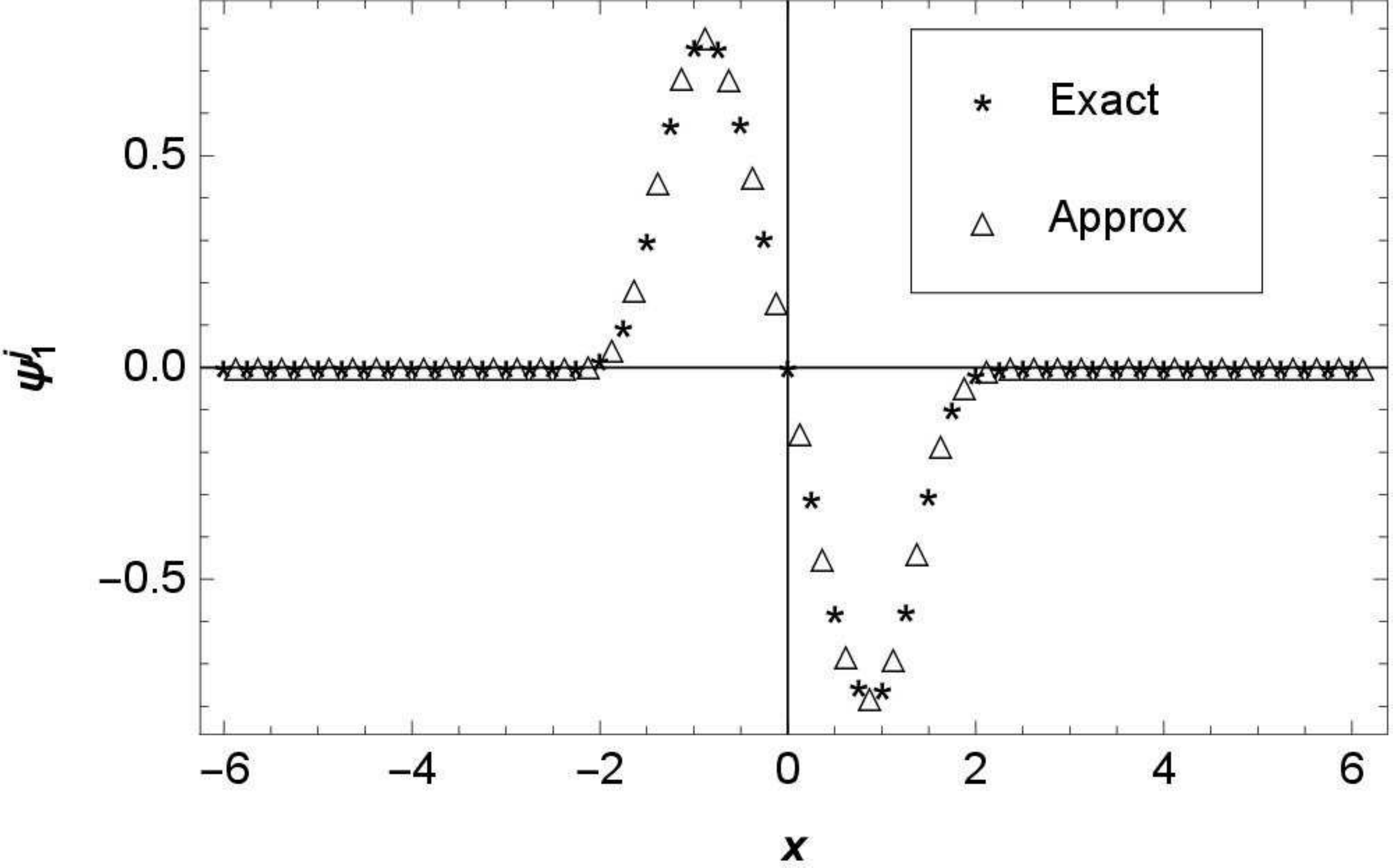}}
		\caption{The same as Figure~\ref{fig1} 
			for the  first excited state eigenfunction ($\Psi_1$)  with $a=-\frac{19}{4},\ b=1,\ c=1$.} 
		\label{fig2}
	\end{minipage}\qquad
    \begin{minipage}{.4\textwidth}
    	\centering
    	\resizebox{0.85\textwidth}{!}{\includegraphics{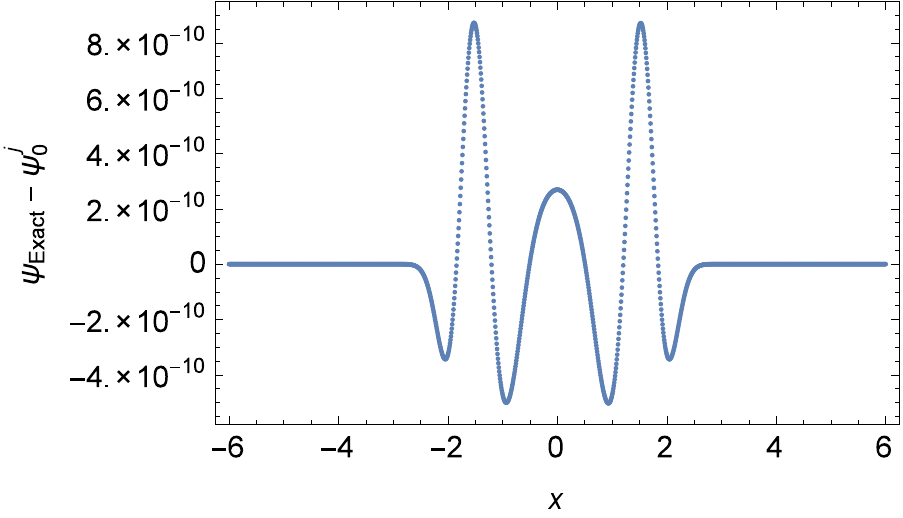}}
    	\caption{Absolute deviation of the ground state wavefunctions ($\Psi_0$) presented in Figure~\ref{fig1}.}
    	\label{fig3}
    \end{minipage}\qquad
    \begin{minipage}{.4\textwidth}
    	\centering
    	\resizebox{0.85\textwidth}{!}{\includegraphics{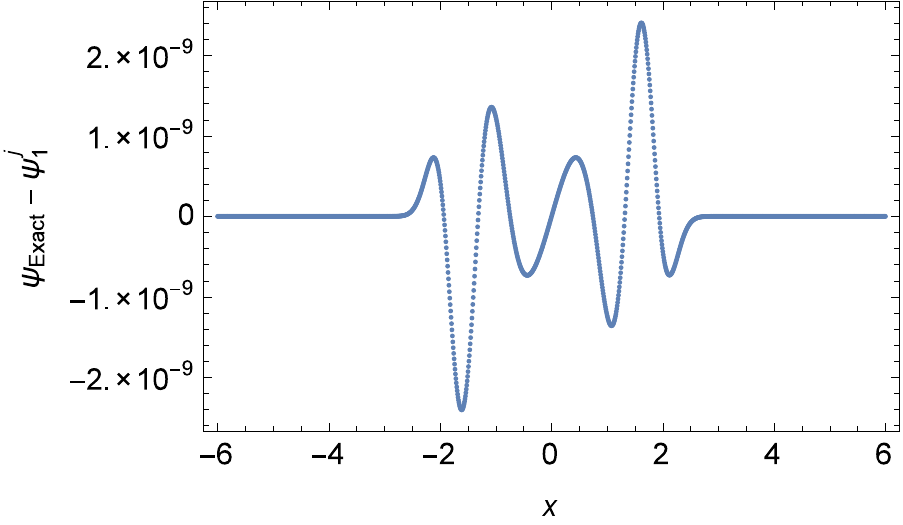}}
    	\caption{Absolute deviation of the of first excited states ($\Psi_1$)  displayed in Figure~\ref{fig2}. }
    	\label{fig4}
    \end{minipage}
\end{figure}
Wavefunction contains all information about the system and hence it is very important to compute it to understand the system. We plot the normalize ground state (Figure~\ref{fig1}) and  first excited state (Figure~\ref{fig2}) wave function obtained from the exact calculation \cite{maiz2018sextic} and DDIPSF (for $j=7$) for the potential eq.(\ref{poten6}) with the constraint on the parameters $a=\frac{b^2}{4c}-\alpha \sqrt{c}$ taking $b=c=1,\ \alpha=3$ and $b=c=1,\ \alpha=5$, respectively. It is to be noted from Figure~\ref{fig1} and Figure~\ref{fig2} that the ground state, as well as the first excited state wavefunctions obtained from DDIPSF method, agrees extremely well with the exact solutions  given by \cite{maiz2018sextic}. As expected, it is seen from the Figure~\ref{fig3} that  the maximum deviation of DDIPSF wavefunction of the  ground state with respect to the exact wavefunction is very small (of the order of $10^{-10}$). Similar, accuracy in the wavefunction of first excited state is noticed from Figure~\ref{fig4}.
We also compare our result for the sextic potential with those obtained by Chaudhuri \textit{et al.}\cite{chaudhuri1991improved} using improved Hill determinant method ($45\times 45$) in Table 5  taking the potential parameters ($a=1,b=-4,c=1$) and ($a=4,b=-6,c=1$) for which the exact energy eigenvalues and eigenfunctions for  the ground state ($E_0$) and first excited state ($E_1$) are  available. Much improved accuracies (absolute error$={E^{exact}-E^{approx}\approx 10^{-11}}$) are achieved from DDIPSF in comparison to those given by \cite{chaudhuri1991improved}. In a recent article, Gaudreau {\it et. al.} \cite{gaudreau2015computing}, calculated energy eigenvalues of the same states using the SCM with double exponential transformation. The solutions (energy eigenvalues) are shown to get converged with an absolute error $\approx 10^{-12}$ which are similar to the solutions from DDIPSF with $j=7$,and  $N=4$.
\begin{table}[ht]
	\begin{center}
		\caption{Comparison of the absolute error of ground state and first excited state energy levels from DDIPSF for the potential eq.(\ref{poten6}) with those from Hill determinant method\cite{chaudhuri1991improved}. Exact values are taken from\cite{banerjee1978k}\label{chousixev}}.
		\begin{tabular}{|c|c|c|c|c|c|c|}
			\hline
			$E_n$& $a$ & $b$ & $c$ & Available  &  Absolute error & Absolute error \\
			& & & & Exact value in \cite{banerjee1978k} & in \cite{chaudhuri1991improved}  &  for $j=7$\\
			\hline
			$E_0$ & $1$ & $-4$ & $1$ & $-2$ & $1(-6)$ & $1.13(-11)$  \\
			\hline
			$E_1$ &$4$ & $-6$ & $1$ & $-9$ & $1.24(-3)$ & $2.16(-11)$\\ 
			\hline
		\end{tabular}
	\end{center} 
\end{table} 
\begin{figure}[h]
	\centering
	\begin{minipage}{.46\textwidth}
		\centering
		\resizebox{0.85\textwidth}{!}{\includegraphics{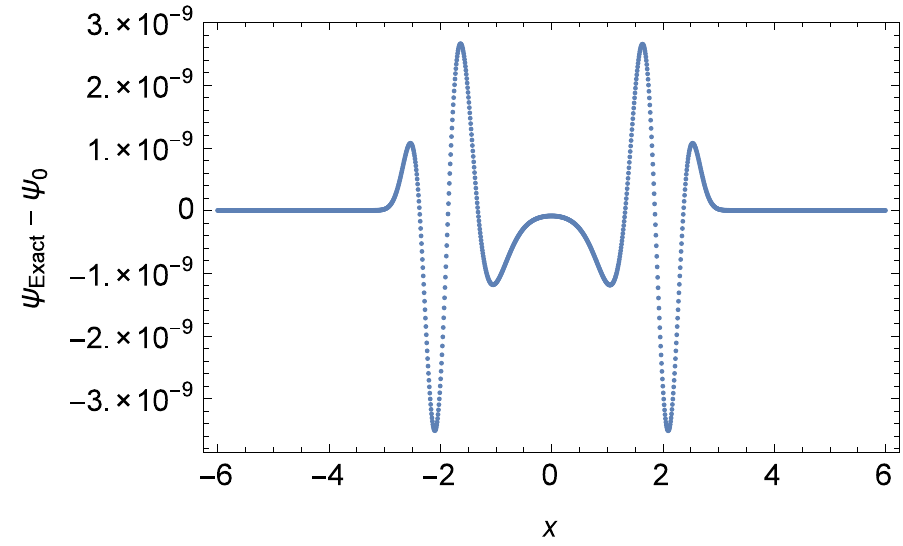}}
	\end{minipage}\qquad
	\begin{minipage}{.46\textwidth}
		\centering
		\resizebox{0.85\textwidth}{!}{\includegraphics{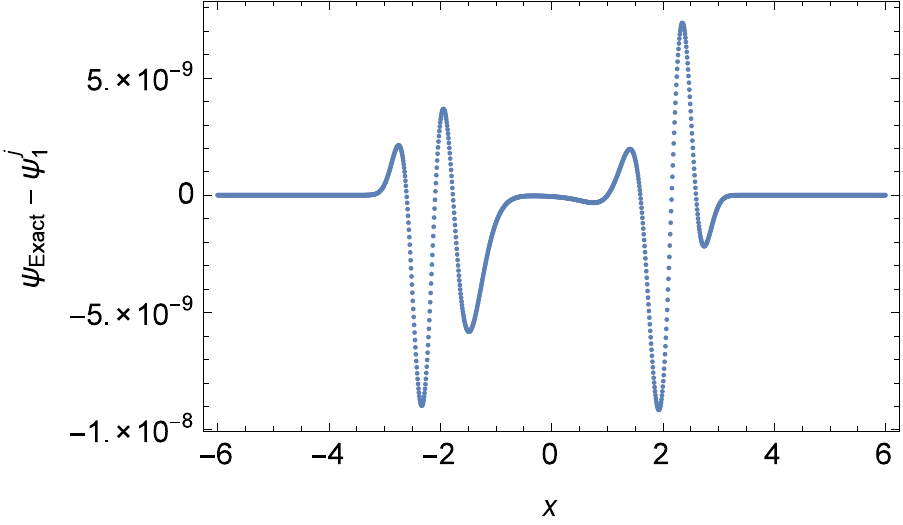}}
	\end{minipage}
	\bigskip
	\begin{minipage}[t]{.46\textwidth}
		\centering
		\caption{Absolute error in DDIPSF normalize ground state eigenfuction with respect to the extact one \cite{chaudhuri1991improved} for the potential eq.(\ref{poten6}) with $a=1,\ b=-4,\ c=1$.}
		\label{choufiga}
	\end{minipage}\qquad
	\begin{minipage}[t]{.46\textwidth}
		\centering
		\caption{Absolute error in DDIPSF normalize first excited state eigenfuction with respect to the extact one \cite{chaudhuri1991improved} for the potential eq.(\ref{poten6}) with $a=4,\ b=-6,\ c=1$.}
		\label{choufigb}
	\end{minipage}
\end{figure}
We have computed the wavefunctions, $\Psi_0$ and $\Psi_1$ for the sextic potential with the values of the parameters given in Ref. \cite{chaudhuri1991improved} ($a=1,\ b=-4,\ c=1$) and resolution ($j=7$) for the purpose of comparison with the wave function given by  Chaudhuri {\it et. al}. We find the maximum deviation of the solution obtained by the present method from the exact one is of the order of $10^{-9}$ as shown in Figure~\ref{choufiga} and Figure~\ref{choufigb}.

It is found that DDIPSF provides highly accurate energy eigenvalues and eigenfunctions for the exactly solvable states of the sextic potential. It is interesting to apply DDIPSF for finding solutions for non-exactly solvable states of the sextic potential. 
In Table 6, we present energy eigenvalues from DDIPSF method ($j=7$) and compare with those obtained by Gaudreau {\it et. al.} \cite{gaudreau2015computing}, using the SCM  for the arbitrary values of the potential parameters. 
\begin{table}[ht]
	\begin{center}
	\caption{The ground state energy values for different potential parameters for the sextic potential in eq.(\ref{poten6}) are obtained in the method based on DDIPSF and compared with those from \cite{gaudreau2015computing} }\label{sinc6}
		\begin{tabular}{|c|c|c|c|c|c|c|}
			\hline
			Parameter & $a$ & $b$ & $c$ & Method in \cite{gaudreau2015computing} & Present method for $j=7$& Difference \\
			Set & & & &($E_{SCM}$)&($E_{DDIPSF}^7$)&($E_{SCM}-E_{DDIPSF}^7$)\\
			\hline
			1&	$0.1$ & $0.1$ & $0.1$ & $0.7646953149964302$ & $0.7646953150098192$ &1.34(-11)  \\ \hline
			2&	$1$ & $1$ & $1$ & $1.6148940820343036$ &  $1.6148940820258968$ & 8.4(-12)  \\ \hline
			3&	$0.1$ & $1$ & $10$ & $2.1277742176946535$ & $2.1277742177121657$ & 1.75(-11)  \\ \hline
			4&	$1$ & $10$ & $10$ & $2.7940871778594101$ & $2.794087177848035$ & 1.14(-11) \\ \hline
			5&	$10$ & $10$ & $10$ & $3.8948206179865981$ & $3.8948206179694673$ & 1.71(-11) \\ \hline
			6&	$-0.1$ & $0.1$ & $0.1$ & $0.663830172742079$ & $0.6638301727525673$ &1.05(-11) \\ \hline
			7&	$1$ & $-1$ & $1$ & $1.2022669303165900$ &  $1.202266930326016$  & 0.94(-11)\\  \hline
			8&	$-0.1$ & $-1$ & $10$ & $1.9385567907196897$ & $1.938556790725087$ & 5.4(-12)  \\ \hline
			9&	$-1$ & $10$ & $10$ & $2.5157308558338656$ & $2.5157308558402014$  & 6.3(-12)\\ \hline
			10&	$10$ & $-10$ & $10$ & $2.9588710692969618$ & $2.9588710692978175$  & 8.5(-13)\\ \hline
		\end{tabular}
	\end{center}
\end{table}
It is found that our results for the energy eigenvalues compare very well (difference $\approx 10^{-11}$) with the results obtained from SCM. 

It was noted in Table 4 that the accuracy increases with higher $j$. To look into another aspect, i.e. the convergence of the energy eigenvalues with respect to 
  $j$, we calculate the energy differences  ($\delta E_{j+1}=E_{DDIPSF}^j-E_{DDIPSF}^{j+1}$) of the ground state of the sextic potential potential with the parameters sets taken in the Table 6. It is observed from the Table 7  that $\delta E_j$s are  decreasing with increasing $j$ which shows a clear trend of convergence of energy eigenvalues with respect to $j$ for all parameter sets.   
\begin{table}[ht]
	\begin{center}
		\caption{The effect $j$ on the ground state energy values for the   parameters sets considered in Table 6.}\label{sinc6a}
		\begin{tabular}{|c|c|c|c|c|}
			\hline
			Parameter &   &   & & \\
			Set & $\delta E_4$ & $\delta E_5$ &  $\delta E_6$& $\delta E_7$\\
			\hline
			1& 1.1(-7)& 1.8(-9) & 2.8(-11)	  & 1.4(-11)    \\ \hline
			2&1.1(-6) & 1.8(-8) & 3.0(-10)	  & 6.1(-12)    \\ \hline
			3&1.0(-5) & 1.8(-7) & 2.8(-9)	  & 2.9(-11)   \\ \hline
			4&1.4(-5) & 2.3(-7) & 3.7(-9)	  & 6.7(-11) \\ \hline
			5&1.8(-5) & 3.0(-7) & 4.7(-9)	 &9.9(-11)   \\ \hline
			6& 1.3(-7) & 2.1(-9) & 2.8(-11)	  &6.0(-12)   \\ \hline
			7&7.7(-7) & 1.3(-8) & 2.0(-10)	 & 4.8(-12)    \\  \hline
			8&1.0(-5) & 1.8(-7) & 2.8(-9)	  & 3.7(-11)   \\ \hline
			9&1.5(-5) & 2.5(-7) &4.0(-9)  &6.4(-11)   \\ \hline
			10&5.6(-6) & 9.2(-8) & 1.5(-9)  & 2.8(-11)  \\ \hline
		\end{tabular}
	\end{center}
\end{table}
It may be interesting to calculate the  wavefunctions corresponding to the non-exactly solvale states. The ground state wavefunctions of  sextic potential for two nonexactly  solvable states are presented in Figure~\ref{fig77} (for low values of the coupling parameters $a=b=c=0.1$ ) and Figure~\ref{fig88} (for high values of the coupling parameters, $a=b=c=10$ ) as an illustration. The form of the wavefunctions are seen to be regular.   
 
\begin{figure}[h]
	\centering
	\begin{minipage}{.46\textwidth}
		\centering
		\resizebox{0.85\textwidth}{!}{\includegraphics{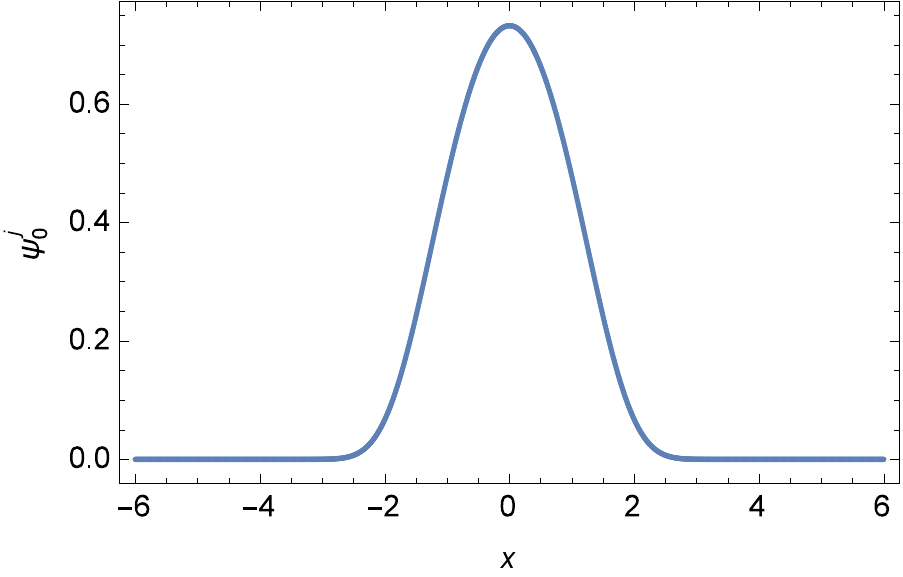}}
	\end{minipage}\qquad
	\begin{minipage}{.46\textwidth}
		\centering
		\resizebox{0.85\textwidth}{!}{\includegraphics{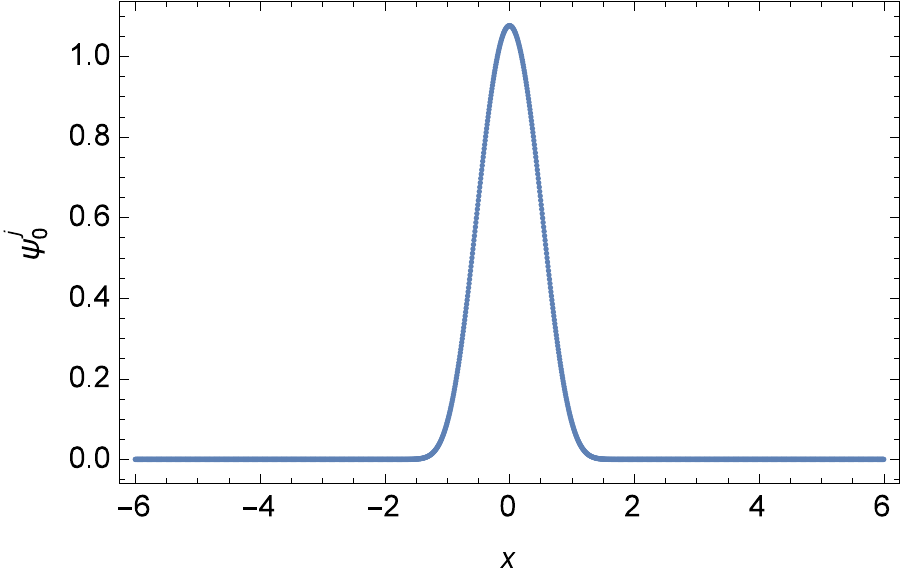}}
	\end{minipage}
	\bigskip
	\begin{minipage}[t]{.46\textwidth}
		\centering
		\caption{Approximate Ground state eigen function for $j=7$ for the potential eq.(\ref{poten6}) with $a=0.1$, $b=0.1$, $c=0.1$.}
		\label{fig77}
	\end{minipage}\qquad
	\begin{minipage}[t]{.46\textwidth}
		\centering
		\caption{Approximate Ground state eigen function for $j=7$ for the potential eq.(\ref{poten6}) with $a=10$, $b=10$, $c=10$.}
		\label{fig88}
	\end{minipage}
\end{figure}

\subsection{Decatic potential}
The decatic AHO, symmetric about the origin, is given by the potential,
\begin{equation}\label{poten10}
V(x)=ax^2+bx^4+cx^6+dx^8+ex^{10}.
\end{equation} 
We have considered the same values of the parameters as given in \cite{gaudreau2015computing}  to compute the energy eigenvalues in the DDIPSF method. It is obvious from the last column of Table 8 that the maximum deviation of the energy eigenvalues between the two methods is mostly $10^{-11}$ for the entire parameter set considered.
\begin{table}[ht]
	\begin{center}
	\caption{Comparison of the ground state energy values  for decatic potential eq.(\ref{poten10}) obtained from DDIPSF (for $j=7$) with those given in \cite{gaudreau2015computing} }\label{dectab}
		\begin{tabular}{|c|c|c|c|c|c|c|}
			\hline
			$a$ & $b$ & $c$ & $d$ & $e$  &$E_{DDIPSF}$ &Deviation  \\	
			$ $ & $ $ & $ $ & $ $ & $ $  &(for $j=7$)& ($E_{SCM}-E_{DDIPSF}$)   \\	
			\hline
			$0.1$ & $0.1$ & $0.1$ & $0.1$ & $0.1$ &  $1.0520482473140977$&$1.53(-11)$  \\	\hline
			$0.1$ & $0.1$ & $1$ & $1$ & $1$ &  $1.5773348519277883$&$6.50(-11)$  \\	\hline
			$1$ & $1$ & $1$ & $10$ & $10$ &  $2.423730003171403$&$1.31(-10)$   \\	\hline
			$1$ & $10$ & $10$ & $10$ & $10$ &  $3.02754208944967$ &$1.83(-10)$  \\	\hline
			$10$ & $10$ & $10$ & $10$ & $10$ &  $4.032920286641719$ & $3.96(-11)$   \\	\hline
			$-0.1$ & $-0.1$ & $0.1$ & $0.1$ & $0.1$ &  $0.9256239552224463$&  $1.97(-11)$ \\	\hline
			$0.1$ & $0.1$ & $-1$ & $-1$ & $1$ &  $0.8618745526018377$&$3.67(-11)$  \\	\hline
			$-1$ & $1$ & $1$ & $-10$ & $10$ &  $1.3353894630249523$&$1.28(-10)$  \\	\hline
			$1$ & $-10$ & $-10$ & $10$ & $10$ &  $1.0275704201159295$& $1.30(-11)$  \\	\hline
			$-10$ & $-10$ & $-10$ & $-10$ & $10$ &  $-22.446238128183488$&$1.61(-9)$  \\	\hline
		\end{tabular}
	\end{center}
\end{table}
We calculate the energy eigenfunctions for the ground state with potential parameters $a=\frac{105}{64}$, $b=-\frac{43}{8}$, $c=1$, $d=-1$, $e=1$ as well as the first excited state with potential parameters $a=\frac{169}{64}$, $b=-\frac{59}{8}$, $c=1$, $d=-1$, $e=1$ for the decatic potential eq.(\ref{poten10}) by employing DDIPSF method with resolution $j=7$ and plotted in Figure~\ref{decfig1} and Figure~\ref{decfig2}.
\begin{figure}[h]
	\centering
	\begin{minipage}{.46\textwidth}
		\centering
		\resizebox{0.85\textwidth}{!}{\includegraphics{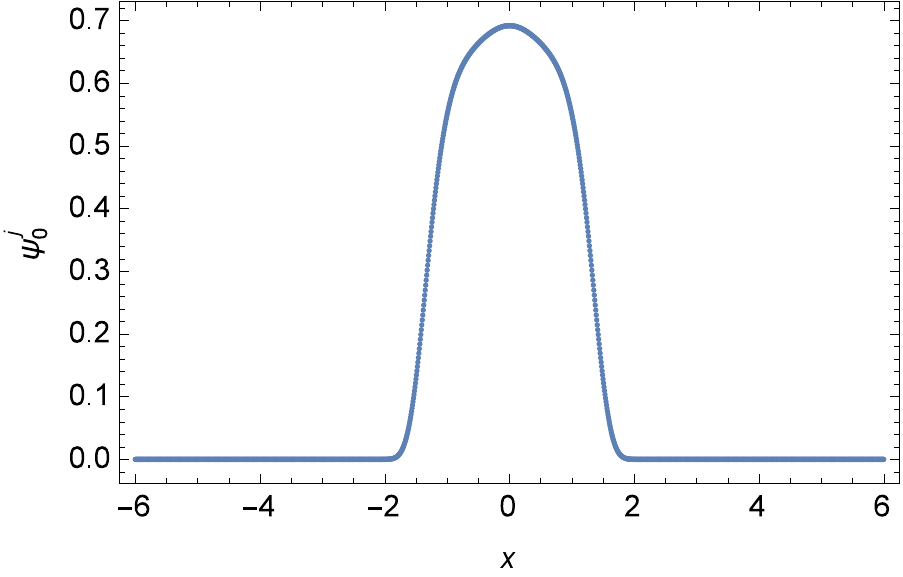}}
	\end{minipage}\qquad
	\begin{minipage}{.46\textwidth}
		\centering
		\resizebox{0.85\textwidth}{!}{\includegraphics{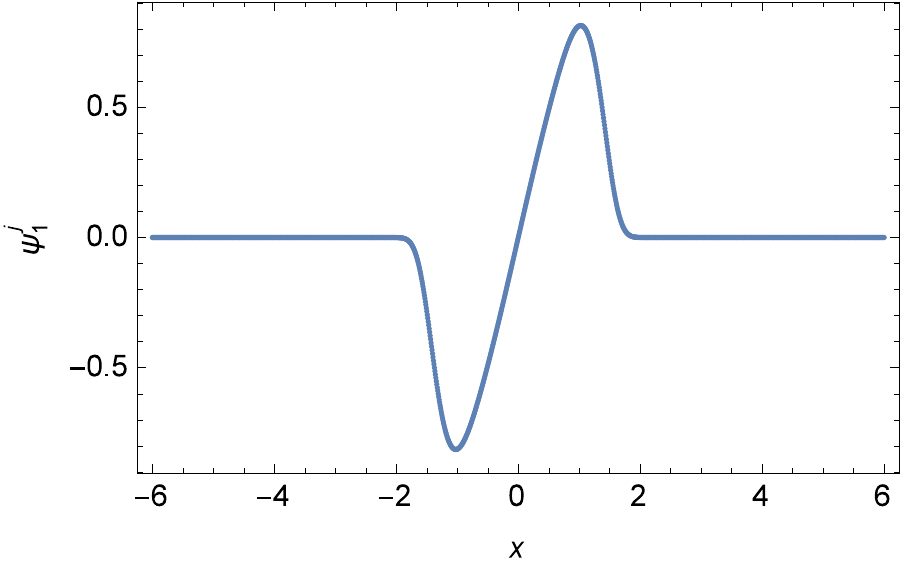}}
	\end{minipage}
	\bigskip
	\begin{minipage}[t]{.46\textwidth}
		\centering
		\caption{Plot of the ground State wavefunction $\Psi_0$  of the potential eq.(\ref{poten10}) with $a=\frac{105}{64}$, $b=-\frac{43}{8}$, $c=1$, $d=-1$, $e=1$ using DDIPSF method for $j=7$.}
		\label{decfig1}
	\end{minipage}\qquad
	\begin{minipage}[t]{.46\textwidth}
		\centering
		\caption{Plot of the first excited state wavefunction $\Psi_1$  of the potential eq.(\ref{poten10}) with $a=\frac{169}{64}$, $b=-\frac{59}{8}$, $c=1$, $d=-1$, $e=1$ using DDIPSF method for $j=7$.}
		\label{decfig2}
	\end{minipage}
\end{figure}

\begin{figure}[h]
	\centering
	\begin{minipage}{.46\textwidth}
		\centering
		\resizebox{0.85\textwidth}{!}{\includegraphics{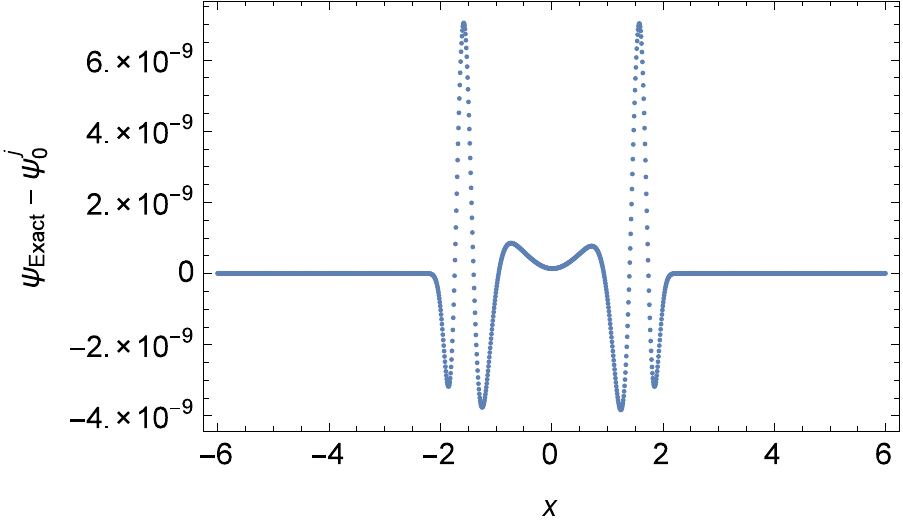}}
	\end{minipage}\qquad
	\begin{minipage}{.46\textwidth}
		\centering
		\resizebox{0.85\textwidth}{!}{\includegraphics{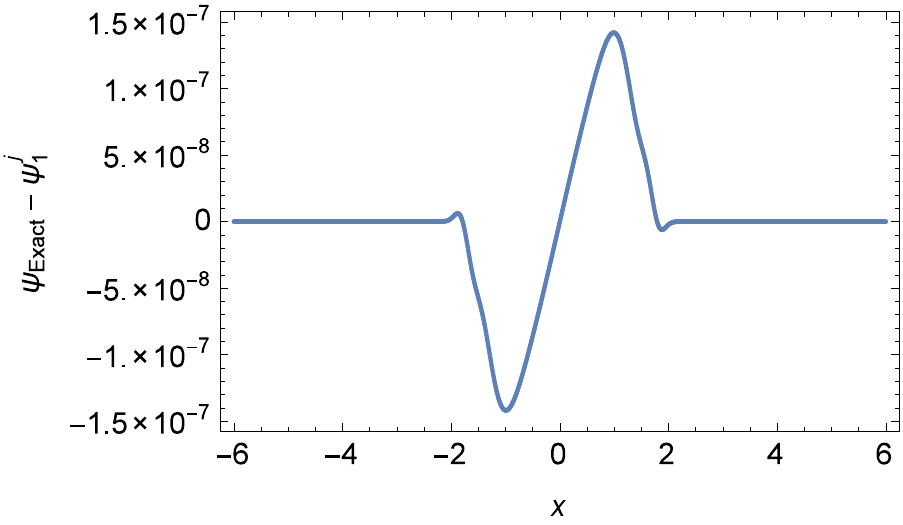}}
	\end{minipage}
	\bigskip
	\begin{minipage}[t]{.46\textwidth}
		\centering
		\caption{Absolute error in DDIPSF wavefunction (for $j=7$) of the ground state for the potential eq.(\ref{poten10}) with $a=\frac{105}{64}$, $b=-\frac{43}{8}$, $c=1$, $d=-1$, $e=1$ with respect to\cite{chaudhuri1991improved}.}
		\label{decfig3}
	\end{minipage}\qquad
	\begin{minipage}[t]{.46\textwidth}
		\centering
		\caption{Absolute error in DDIPSF wavefunction (for $j=7$) of the first excited state for the potential eq.(\ref{poten10}) with $a=\frac{169}{64}$, $b=-\frac{59}{8}$, $c=1$, $d=-1$, $e=1$ with respect to\cite{chaudhuri1991improved}.}
		\label{decfig4}
	\end{minipage}
\end{figure}
As the wavefunction is not available in \cite{gaudreau2015computing}, we have taken the exact wavefunction given in \cite{chaudhuri1991improved} for the purpose of comparison. Figure~\ref{decfig3} and Figure~\ref{decfig4}, display the errors of our calculation with respect to the corresponding exact values taking the resolution, $j=7$ and potential parameters $a=\frac{105}{64}$, $b=-\frac{43}{8}$, $c=1$, $d=-1$, $e=1$ and $a=\frac{169}{64}$, $b=-\frac{59}{8}$, $c=1$, $d=-1$, $e=1$ respectively. It is found that for the entire range of $x$, a maximum error of $10^{-9}$ and $10^{-7}$ for the ground state and first excited state, respectively. So, DDIPSF gives very accurate results for all cases considered.

\section{Conclusion}
We have applied DDIPSF method to study the sextic and decatic AHOs. It is seen from the recent literature that the exact solution for such AHOs can be found only for selective states for particular choices of the values of the potential parameters \cite{maiz2018sextic,brandon2013exact}. This method gives values of energies of eigenstates with an accuracy up to the twelveth decimal  for the resolution $j=7$ and $N=4$. Eigenfunctions, obtained by the present method mimics very well with the available exact values for the entire domain of $x$ with a maximum error of  ${O}(10^{-9})$.  DDIPSF method is very simple and very fast to converge to yield very accurate results. The advantages of the present method is the ability to find both the energy eigenvalues and eigenfunctions with high accuracy for arbitrary choices of the potential parameters. 


%
%

\end{document}